\documentclass[conference]{IEEEtran}
\usepackage{blindtext, graphicx}

\renewcommand{\baselinestretch}{1.05}

\ifCLASSINFOpdf
\else
\fi
\begin{document}
%

\title{Derivation of Network Reprogramming Protocol with Z3}

\author{\IEEEauthorblockN{Vidhya Tekken Valapil}
\IEEEauthorblockA{Department of Computer Science\\
Michigan State University\\
East Lansing, Michigan 48824\\
Email: tekkenva@cse.msu.edu}
\and
\IEEEauthorblockN{Sandeep S. Kulkarni}
\IEEEauthorblockA{Department of Computer Science\\
Michigan State University\\
East Lansing, Michigan 48824\\
Email: sandeep@cse.msu.edu}
}


%


\maketitle



%

\label{sec:Abstract}
\section{Abstract}


Networks protocols are the heart of communication networks. An efficient network protocol does maximum utilization of the underlying network capabilities.
Network Protocol synthesis is the process of synthesizing or deriving network specific protocols from the requirements of a given specific network. In this report, we present a step-by-step approach for the automated synthesis of network protocols from the network specifications. Using SMT solvers to automate the protocol generation is the key idea behind the presented synthesis approach. The protocols generated using this approach followed the most optimal way of data transmission for the given network requirements.

\label{sec:Introduction}
\section{Introduction}



Communication in any given network is guided and controlled by the protocol specified for that network, which are a set of rules, procedures and formats that  any device in that network is required to comply. 
In other words, every network has its own set of requirements, and at any point in time every component of the network follows or acts as per a specific manual that helps satisfy these requirements. 
Given the network requirements, the goal of this project is to automate the generation of such a manual from those requirements. 
This process of generating or synthesizing the network protocol from the network requirements is called protocol synthesis.
With several existing network protocols already available, generating an entirely new protocol for a network may seem inessential. 
However, generating or synthesizing network specific protocols has its own advantages. 
Especially, the problem with existing protocols is that they focus on optimizing only one or two resources. For example, protocols like Ad hoc On-Demand Distance Vector \cite{AODV}, Dynamic Source Routing in Ad Hoc Wireless Networks \cite{DSR}, Adaptive Distance Vector \cite{ADV}, Destination-sequenced distance-vector routing \cite{DSDV}, DREAM \cite{DREAM} (Position Based Routing), LAR \cite{LAR} (Position Based Routing), etc., focus on performing routing in such a way that routes span minimal distance or are computed dynamically based on the current location of destination node. However, when communicating between two devices distance is not the only requirement that needs to be minimized. There are several other features like data rate, bandwidth, power, loss rate, etc., that may have to be considered as the requirements of a given network. So using existing protocols may not be the best solution when networks have wide variety of requirements. This calls for the need for customized network protocols that take into consideration several types of network requirements.

\begin{figure}
  \centering
  \includegraphics[width = 0.4 \textwidth]{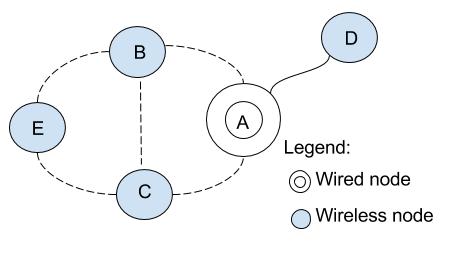}
  \label{fig:network_example_1}
  \caption{Hybrid Network Example}
\end{figure}

For instance consider the simple real world scenario shown in figure \ref{fig:network_example_1}. $A$, $B$, $C$, $D$, $E$ are nodes in the given home/office network. Nodes $B$, $C$, $E$ are connected wireless to the other devices in the network, denoted by dotted lines, and nodes $A$ and $D$ are connected wirelessly. 
Node $A$ has unlimited power, whereas nodes $B$ and $C$ run on battery. 
Node $B$ is closer to node $C$ than node $A$, and similarly the distance between nodes $C$ and $A$ is more than the distance between nodes $C$ and $B$. 
Now in such a network, if one chooses to use distance based protocols like AODV, ADV or DSDV, route $B$ to $C$ would be preferred for data transmission than the route or link between nodes $B$ and $A$ since $B$ is closer to $C$ than to $A$. 
However, since $A$ has higher power capability it would be better for $B$ to acquire/transmit data to $A$ rather than $C$ since $C$ has limited power which could be saved or utilized for several other critical tasks. One possible solution to this problem could be the use of energy efficient network protocols like \cite{EnergyForPLC}, \cite{EnergyForAdhoc}, \cite{DistrPwCtrlForAdhoc}. But problem lies in the fact that distance and power are not the only requirements that have to be considered in any given network. For example in the same network let node $B$ be a security device that performs intruder sensing and triggers an alarm when needed, and let node $C$ be a video monitoring device. In this scenario the requirement of node $B$ is low latency where as with node $C$ the requirement is high bandwidth and low jitter. So a network protocol that is tailored to the various aspects of this network is required. Moreover, customized protocols  achieve maximum utilization of the underlying network capabilities, which is not guaranteed by the existing general purpose protocols.

\begin{figure}
  \centering
  \includegraphics[width = 0.4 \textwidth]{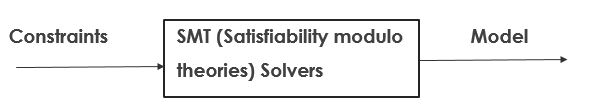}
  \label{fig:smt_solver}
  \caption{SMT Solver outline}
\end{figure}

Satisfiability Modulo Theorem (SMT) Solvers \cite{SMTSolvers} have proven to be a blessing to mathematicians and computer scientists in the past few decades. 
They are useful for automated verification of programs, proving correctness of programs, symbolic execution based software testing, and for synthesis \cite{SMTwiki}.
Given a set of logical formulas as input, SMT solvers answer the decision problem that if the given set of logical formulas (that may share common variables) are satisfiable, i.e., if there exists some set of values for the variables included in the formulas such that the logical formulas always evaluate to true.
If the provided logical formulas are satisfiable, then the solver provides values (or range of values) that satisfy the given formulas.
If the provided logical formulas are not satisfiable, then the solver provides counter examples or indicates aspects that refrain the formulas from being satisfied.
These input formulas are also called assertions or constraints that need to hold at all times. The satisfying values returned by the solver when the given formulas are satisfiable are also called models.

\begin{figure}
  \centering
  \includegraphics[width = 0.4 \textwidth]{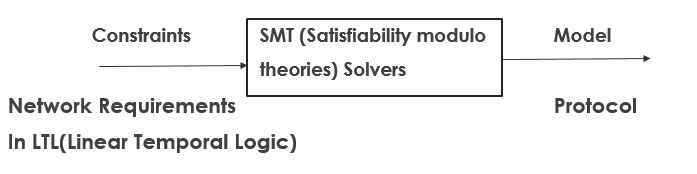}
  \label{fig:smt_solver_network}
  \caption{SMT Solver with network requirements as input}
\end{figure}

Z3 \cite{Z3} is one of the most widely used SMT solvers developed at Microsoft Research and has been an essential part of several research works in cloud computing \cite{Z3inAzure}, networking \cite{Z3NtwkConnPolicies},\cite{checking-beliefs-in-dynamic-networks}, model checking, etc. As mentioned in the previous paragraph SMT solvers are used for synthesis, and since Z3 has proven to be useful in  research in networking, we use Z3 in this project to automate the process of protocol synthesis. The assertions or logical formulas provided as input to the SMT Solver Z3 are the network requirements as shown in \ref{fig:smt_solver_network}. If the provided network requirements are satisfiable then Z3 provides a model which corresponds to the desired network protocol. By following this approach network protocols for broadcasting in shared channel network, and for data transmission in a simple neighborhood based network were generated. The generated data transmission protocols were simulated in NS3 and compared with slightly modified versions of an existing wireless protocol in NS3 with wifi nodes based on CSMA to evaluate its performance.

\textbf{Organization of the paper.} The rest of the report is organized as follows: in section \ref{sec:Related Work} we will discuss some of the existing research performed on the topic of protocol synthesis. In section \ref{sec:Methodology or Technical Approach} we briefly introduce the methods used in the project for synthesizing network protocols. In section \ref{sec:Design Rationale} we will provide further design details of the method used for synthesis. In section \ref{sec:Technical Results and Analysis} we discuss the results obtained such as the network protocols generated and their simulation results in the network simulator. In section \ref{sec:Future Work} we consider some extended research ideas and methods to overcome the shortcomings of the currently used methods and methods to widen their applicability. Finally, in \ref{sec:Summary and Lessons Learned} we provide the overall project summary and state some lessons learnt during the course of this project.

\label{sec:Related Work}
\section{Related Work}


\begin{figure}
  \centering
  \includegraphics[width = 0.4 \textwidth]{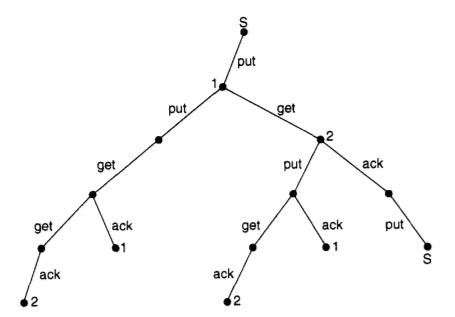}
  \label{fig:ordering_tree_example}
  \caption{Ordering Tree Example \cite{AutoProtSynthCarchiolo}}
\end{figure}

\begin{figure}
  \centering
  \includegraphics[width = 0.4 \textwidth]{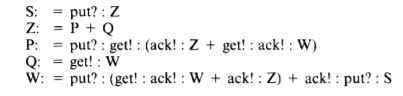}
  \label{fig:CCS_notation}
  \caption{Corresponding CCS notation \cite{AutoProtSynthCarchiolo}}
\end{figure}

In \cite{AutoProtSynthCarchiolo} Carchiolo et.al perform synthesis of protocols for co-operative systems starting from formal description of communication requirements. They encode the network requirements from the user in temporal logic discussed in section \ref{sec:Design Rationale}. 
These requirements are then mapped to ordering trees. A sample temporal ordering tree is shown in figure \ref{fig:ordering_tree_example} from \cite{AutoProtSynthCarchiolo}. 
These are then further mapped to behavioral specifications expressed in Calculus of communicating systems (CCS). CCS notation corresponding to figure \ref{fig:ordering_tree_example} is shown in figure \ref{fig:CCS_notation}. 
Apart from these specifications, CCS equations for missing modules are created and solved to acquire specifications for missing modules also in CCS using the the synthesis method used in \cite{CCSmappingGiordano}. However, the process of creating these CCS equations is not automated and requires expertise to analyze and express the missing internal requirements. Thus a completely automated synthesis procedure without any required intervention or need for intermediate analysis during the synthesis process is desirable.

\begin{figure}
  \centering
  \includegraphics[width = 0.4 \textwidth]{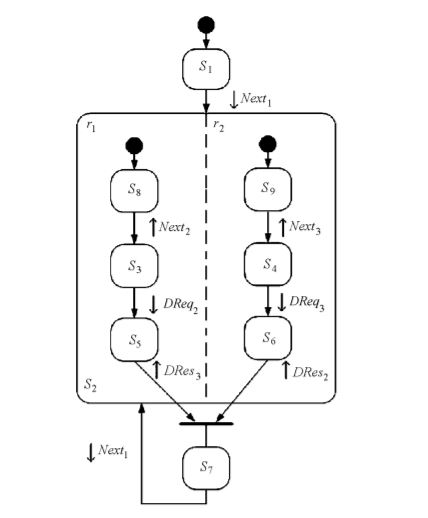}
  \label{fig:service_spec_UML}
  \caption{An example of service specification in the form of UML state machine \cite{AutoProtSynthUMLSpec}}
\end{figure}

In \cite{AutoProtSynthFSMSpec} Saleh K. et.al performed communication protocol synthesis from communication service specifications using modeling in finite state machines\cite{FSM}. 
The service specifications are expressed as finite state machines, 
they are then projected onto the internal set of service access points in the system. 
Now, a set of conditions are evaluated against the transitions in the projection obtained in the previous step. 
These conditions involve service states and service access points. 
Based on the result of the evaluation of each transition a particular transition is synthesized. The resulting set of synthesized transitions form the set of protocol traces which are then further optimized to obtain the final protocol which are are still in the form of finite state machines. The problem with this protocol synthesis approach is the expressiveness of finite state machines. Finite State machines are unable to express concurrent service behaviors, and concurrency is one of the critical aspects of communication in any given network. 
However in one of their recent works \cite{AutoProtSynthUMLSpec} Saleh K. et. al. overcame this problem by using UML state machines that have better expressiveness than finite state machines. Specifically, UML protocol state machines can be used to express concurrency and state hierarchy, consider figure \ref{fig:service_spec_UML} from \cite{AutoProtSynthUMLSpec}. So in \cite{AutoProtSynthUMLSpec} the service specifications are modeled as UML protocol state machines and they apply similar projection and evaluation steps as in \cite{AutoProtSynthFSMSpec} to transform the service specifications to protocol specifications but in the form of UML protocol state machines. However this approach falls short when it comes to synthesizing protocols with timing and security constraints, i.e., it is inapplicable to real applications that have timing constraints or security conditions as part of their requirement. Thus a synthesis approach that is capable of considering different types of network constraints including timing and security constraints is required.

\label{sec:Methodology or Technical Approach}
\section{Methodology or Technical Approach}

 
For any given network, choosing a protocol that achieves the best out of the capabilities of that network depends on finding a protocol that considers all different requirements of that network. 
In other words, only a protocol that considers all aspects or specifications of a given network can attain maximum utilization of capabilities of that network. This necessitates the need for customized protocols for communication networks and the process of generating customized network protocols from requirements or specifications of the given specific network is called protocol synthesis. Though protocol synthesis is desirable existing protocol synthesis approaches have  limitations as discussed in section \ref{sec:Related Work}. So in this project we propose an approach for automated synthesis of network protocol from network specification using SMT solvers that overcomes the discussed limitations of existing approaches. 

The experimental method for automated protocol synthesis used in this project consists of the following steps:

1. For a given specific network, identifying constraints based on problem requirements i.e., requirements of the given network.

2. Encoding requirements or specifications in computational temporal logic.

3. Provide the temporal logic formulas as assertions or problem constraints to the SMT solver.

4. Run the SMT solver and verify if the constraints are satisfiable. 

5. If the constraints are not satisfiable, analyze the counter examples generated by the solver to detect any conflicting requirements. Resolve conflicting requirements and continue from step 3.

6. If the constraints (network requirements) are satisfiable then the SMT solver provides a satisfying model.

7. Extract node or process actions from the generated model and store it onto a model-file for further parsing.

The extracted protocols are then simulated in NS3 and analyzed against a slightly modified version of an existing wireless protocol in NS3 with wifi nodes based on CSMA. This protocol was chosen for comparison due to the fact that it was the closest available protocol applicable to the network scenario and requirements considered in the project. Estimations such as the total amount of data transmitted were computed by manually analyzing the NS3 log.

\label{sec:Design Rationale}
\section{Design Rationale}


In this section we elaborate each of the steps discussed in section \ref{sec:Methodology or Technical Approach} by providing the design details involved in each step. We will walk-through with an example of a network scenario, from requirements specification to acquiring the corresponding protocol.

\textbf{Step 1:}

{\em For a given specific network, identifying constraints based on problem requirements i.e., requirements of the given network.}

Consider a network with the following requirements,

\begin{itemize}
\item{A process at any given time does exactly one of the three actions: transmit or listen or sleep.}

\item{A process in this network at any point in time transmits garbage (can denote exchange of some control information to let neighbor process know that it is alive) or valid packet or does not transmit at all.}

\item{A process performs every action infinitely often. This is to ensure liveness in terms of network actions.}

\item{In the initial state there is exactly one process in the network that has all the information that needs to be transmitted, i.e.exactly one process knows all data packets in the initial state of the network.}

\item{A process can transmit only the data packets that it knows.}

\item{When a process learns about some data packet it never forgets it.}

\item{A process learns about a data packet by listening to a transmitting process when no other process is transmitting (to avoid any collision).}

\end{itemize}

Identifying and gathering network requirements for design is by itself a widely researched topic \cite{ReqElic}, \cite{GatherNetworkReq}. We consider these requirements since they are some of the most common and basic requirements of simple communication networks. In practice network requirements when available in their initial raw form can be more close to natural language i.e. in layman level or can be more sophisticated than those considered here.

\textbf{Step 2:}

{\em Encoding requirements or specifications in computation temporal logic.}

Computational Temporal Logic (CTL) is a widely sorted category of Temporal logic that is useful in formal specification of computer programs or systems, formal analysis, and verification of executions of computer programs and systems. 
The ability to reason about time in CTL is one of the key factors behind its substantial expressiveness. 
Several existing works on automated protocol synthesis use finite state machines and UML state machines. In this project we used Computational Temporal Logic for the formal specification of network requirements.
The advantage of specifying requirements in CTL over other forms of specification like in the form of finite state machines and UML state machines is that with CTL one has the ability to specify constraints involving concurrency and time.
So now we will specify the first requirement mentioned in the previous step formally in Computational Temporal Logic.

{\em "A process at any given time does exactly one of the three actions: transmit or listen or sleep."}

If $t$ stands for time and $p$ represents process then,

$\forall t$ $\{\forall p$ 
$[(p.sleep \rightarrow \neg(p.transmit \wedge p.listen))$ $\vee$

\indent\indent\indent$(p.transmit \rightarrow \neg(p.sleep \wedge p.listen))$ $\vee$

\indent\indent\indent$(p.listen \rightarrow \neg(p.sleep \wedge p.transmit))]\}$

In simpler words the formal specification states that at all times for all processes, when a process transmits it does not sleep or listen, when a process listens it does not sleep or transmit, when a process sleeps it does not transmit or listen. One may observe that this constraint does not consider the case where a process does not transmit or sleep or listen. When such cases are missed it causes non-determinism to become part of the output model and it may result in models with unexpected behaviors. However there are instances where such non-determinism is desirable, especially when there is no solution to a given problem (set of unsatisfiable constraints) but applying heuristics can help reach a solution by exploring non-deterministic paths. So for the example considered in the current discussion the missing case needs to be included, and we do so by adding a separate constraint as follows,

$\forall t$ $(\forall p (sleep(t,p)==True \vee$ 

\hspace{10mm}$transmit(t,p)>=0 \vee listen(t,p)==True)))$

\textbf{Step 3:}

{\em Provide the temporal logic formulas as assertions or problem constraints to the SMT solver.}

The SMT solver that was used in the project for model or protocol generation was Z3. More specifically Z3Py was used to aid in automated generation of multiple Z3 constraints with slight variations.
In Z3 each requirement or constraint is provided as a assert command that needs to hold at all times. Each of these constraints are added to a common solver instance and are checked for satisfaction together.
The fundamental blocks in Z3 are constants and functions. Functions are defined on all values or inputs in the provided range (if any). 
These functions and constants are uninterpreted or free, which means that they do not have any prior interpretation attached to them. To illustrate this let us consider the formal network specification provided in the previous step. Since $sleep$, $transmit$ and $listen$ are process states (i.e. state of a node in the network) whose values (true or false) vary with time we will encode them as uninterpreted functions of the following form:

\noindent$listen = Function('listen', IntSort(), IntSort(),$

\hspace{65mm} $BoolSort())$ 

\noindent$sleep = Function('sleep', IntSort(), IntSort(),$

\hspace{65mm} $BoolSort())$

\noindent$transmit = Function('transmit', IntSort(), IntSort(),$

\hspace{65mm} $IntSort())$

Last argument of these functions denote what they return. 
It can be observed that $listen$ and $sleep$ both return boolean values, i.e., true if the process (denoted by the second integer argument) is listening (sleeping) or not listening (not sleeping) at that specific time (denoted by the first integer argument).
However unlike $sleep$ and $listen$, that depend only on time and process, as specified in Step 1 $transmit$ is associated with the content being transmitted, i.e.,at a given time a specific process either does not transmit or it transmits garbage or a valid packet. The value returned by $transmit$ i.e., the last integer argument can be configured such that it corresponds to an integer that uniquely distinguishes or denotes one of the three cases. Thus the Z3 constraint corresponding to the formal CTL specification derived in the previous step is as follows:

$s.add(Implies(constraint,$

\indent\indent $And(Implies(sleep(t,p)==True,$

\indent\indent\indent\indent\indent $And((listen(t,p)==False),$

\indent\indent\indent\indent\indent\indent\indent $transmit(t,p)==-1)),$
             
\indent\indent\indent     $Implies(transmit(t,p)>=0,$
				
\indent\indent\indent\indent\indent $And((sleep(t,p)==False),$

\indent\indent\indent\indent\indent\indent\indent $(listen(t,p)==False))),$

\indent\indent\indent		$Implies(listen(t,p)==True,$

\indent\indent\indent\indent\indent $And((sleep(t,p)==False),$

\indent\indent\indent\indent\indent\indent\indent $transmit(t,p)==-1)))$

                      $))$

Constraints corresponding to the remaining requirements specified in Step 1 are available in the file titled "constraints\_neighbors\_broadcast.py" in folder "codes/Z3Py/".

\textbf{Step 4:}

{\em Run the SMT solver and verify if the constraints are satisfiable.}

The SMT solver is run to determine if the provided constraints are satisfiable. If the constraints are not satisfiable we proceed to step 5 and if they are satisfiable we proceed to step 6. The time taken by the solver to decide (to return "SAT" or "UNSAT") increases with the increase in the number of constraints with nested quantifiers (for e.g. $\forall x$ $(\exists y$ $(\forall z$ $(condition)))$). 

\textbf{Step 5:}

{\em If the constraints are not satisfiable, analyze the counter examples generated by the solver to detect any conflicting requirements. Resolve conflicting requirements and continue from step 3.}

If the provided constraints are not satisfiable then the SMT solver returns "unsat" with the list of constraints (network requirements) that cannot be satisfied due to some inter-constraint conflict. These set of constraints are also called the "unsat-core" of the model. One can analyze these constraints to prioritize and eliminate any accordingly, or alter any constant or function domain (widen or contract its input or value space) in these constraints such that any existing conflicts can be resolved or avoided by doing so. Then the updated or modified assertions can be re-checked for satisfaction by repeating from Step 3.

\textbf{Step 6:}

{\em If the constraints (network requirements) are satisfiable then the SMT solver provides a satisfying model.}

\begin{figure}
  \centering
  \includegraphics[width = 0.4 \textwidth]{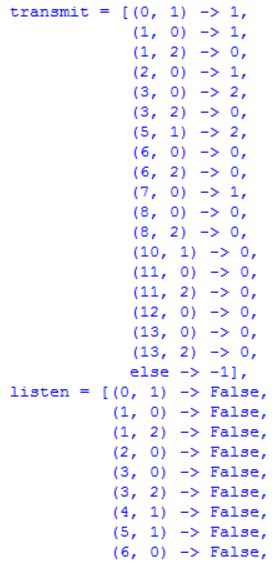}
  \label{fig:model_protocol}
  \caption{A portion of the model returned by Z3}
\end{figure}

If the given set of assertions are satisfiable, then the SMT solver returns "sat" along with the satisfying set of values or arguments for the constants and the functions that are originally uninterpreted in the input assertions.
Remember that the constraints are originally logical formulas that are required to evaluate to true. 
The acquired satisfying set is also called the satisfying model and this satisfying model is nothing but the required network protocol. 
For example, for the requirements considered in this section (specified in step 1), a portion of the model or protocol obtained from Z3 is shown in figure \ref{fig:model_protocol}, where the SMT solver provides a set of values that can assigned as arguments to the uninterpreted functions. The set of values provided by the solver are in fact obtained from the legal domain (allowed range of values) of the constants used in the functions.
This domain could have been specified explicitly as separate constraints or could be the default domain (range of values) of constants considered in the solver internally based on their type (integer, bool, etc.). 

For example as discussed in step 3, $transmit$ is an uninterpreted function that takes two integer arguments (time and process) and returns an integer that denotes the content being transmitted. 
We specified the possible interpretations of the integer value returned by $transmit$ as a separate constraint, i.e. positive values denote valid packets, zero denotes garbage and $-1$ to indicate that the process is not transmitting at that particular time. 
So the first line in the portion of the model shown in the figure \ref{fig:model_protocol} says that at time $t=0$, process (identified by id) $p=1$ transmits packet (identified by id) $1$. Thus one can analyze the acquired model to obtain the protocol traces which correspond to process actions in the network scenario discussed in this section.

\textbf{Step 7:}

{\em Extract node or process actions from the generated model and store it onto a (model-)file for further parsing.}

The returned model is then stored onto a file for extracting specific portions of the protocol, like which process is transmitting what content at each instance of time, when was each process listening to transmissions by other processes and so on. This is done by parsing the file containing the model using some scripting language, python was used in this project. Some portions of the returned model need further processing to understand or interpret them. For example, even for a smaller run i.e. when verifying the constraints against a small set of input values, the returned model can contain portions like $knows = [else -> knows!37(k!36(Var(0)), Var(1), Var(2))]]$ where parts like $knows!37$ are variables created by the model and their values are also provided by the model. These require recursive replacement to make the overall model interpretable.

\label{sec:Technical Results and Analysis}
\section{Technical Results and Analysis}


Since the final protocol is obtained by automated synthesis from the actual network requirements of the underlying network, the protocol is correct by construction, i.e. it should satisfy the requirements of the underlying network. SMT Solvers are in fact used to verify program correctness, they do so by enumerating over the input space and evaluating the program in every possible scenario. In this project we in turn use these solvers to explore all possibilities to first verify if our input requirements are satisfiable and when the solver says that they are satisfiable it does so by finding a way (computing a satisfying set of values) to satisfy or solve the input specifications. It then provides this way as a model to us and we in turn use it as network protocol. Since the approach used in this project does not involve intermediate intervention the obtained protocol is correct by construction and is applicable to the corresponding specific network at hand, as long as the given network requirements were correct.  

We generated two protocols using the approach presented in this report. The first protocol has all the requirements specified in section \ref{sec:Design Rationale} step 1, and additionally we require that by a specific amount of time specified number of packets reach all process, i.e., all processes know all packets eventually after some specific amount of time. We manually change this amount of time incrementally starting from $0$ time units to determine the minimum amount of time required by the processes to complete this transmission. This network did not have any requirement or restriction over which process can listen to which process. 

\textbf{Lemma.}
{\em In such a network the best protocol to transmit the data packets to all processes is by broadcast.}

\textbf{Proof.} Consider a network of $n$ processes where exactly one process say $n_0$ knows $m$ packets at time $0$. If the ultimate goal is that the $n-1$ processes should know the $m$ packets by some finite amount time, then fastest way to achieve this is by making the process $n_0$ to broadcast the packets to all processes simultaneously at the same time one packet after the other. One may doubt that this is not the fastest way for the processes to learn all the packets because the lack of restriction over which process can listen to which process provides a way for processes other than process $n_0$ to also transmit packets as soon as they learn them, thereby reducing the burden on process $n_0$ and quickening the overall transmission. However this solution is not applicable to the considered network scenario because one of the initial requirements states that "A process learns about a data packet by listening to a transmitting process when no other process is transmitting". This clearly eliminates the chance of two process transmitting simultaneously which leaves us with broadcast as the only optimal solution. For any process other than process $n_0$ to transmit a packet, it would take two steps, first the process learns about the packet from process $n_0$ and in the next step it transmits the packet to some other process and no other process transmits during this second step including process $n_0$. Forcing $n_0$ to be idle during any step increases the total time taken to complete the transmission, since process $n_0$ transmits the packets back to back as per the optimal broadcast solution and is never idle, so completes the transmission in $tu$ time units which is equal to the number of packets to be transmitted. 

The protocol or model obtained from Z3 by synthesizing from the network requirements of this network scenario also performs broadcast. Thus our protocol synthesis approach provides the most optimal solution for any given set of network requirements. We simulated the obtained protocol in NS3 to analyze its applicability. We created a wireless topology of three wifi nodes and performed timed broadcast from one node to the others corresponding to the synthesized protocol and compared it with the default non-timed broadcast protocol in a wifi network. 

\begin{table}[h!]
\caption{Broadcast in a 3-node wireless network in 3 seconds}
\centering
 \begin{tabular}{||p{3cm} c||} 
 \hline
 Approach & Bytes Transferred\\ [0.7ex] 
 \hline\hline
 Non-timed broadcast protocol & 187392\\ 
 Synthesized timed broadcast protocol & 184832\\ [1ex] 
 \hline
 \end{tabular}
\end{table}

We can observe that the synthesized time based protocol sends less data than the non-timed CSMA based wifi broadcast protocol. This is because in the timed protocol the node $n_0$ takes some time for starting the application before transmitting the data whereas with the non-timed protocol the data is broadcasted at a stretch and no time is spent on periodic application start. 

The second protocol was based on all network requirements as in the previous network scenario but with additional restriction on which process can listen from which process, i.e., a process can receive only from a process on its immediate left (with process identifier value 1 less than itself) in the linked neighborhood. The synthesized protocol was simulated in NS3 and was compared with the non-timed broadcast protocol in a wifi network with CSMA capability. 

\begin{table}[h!]
\caption{Transmission in line-like topology with 3 wifi nodes in 2 seconds}
\centering
 \begin{tabular}{||p{2cm} c p{2cm} p{1cm}||} 
 \hline
 Approach & Bytes Transferred & CSMA needed for collision detection & Power consumed\\ [0.7ex] 
 \hline\hline
 Non-timed transmission protocol & 55808 bytes & required & $6pw <$\\ 
 Synthesized timed transmission protocol & 55808 & not required & $5pw$\\ [1ex] 
 \hline
 \end{tabular}
\end{table}

We can observe that the synthesized time based protocol sends the same amount of data as the non-timed CSMA based wifi broadcast protocol. 
However with our synthesized protocol collision detection is not required since the processes are timed to transmit at different instances of time resulting in the channel being used exactly by one process at any time.
Let us assume that when a node is up and running (listening or transmitting) every second it utilizes $pw$ units of power. 
Now with our protocol the node $n_0$ spends $pw$ units to transmit to node $n_1$, and simultaneously $n_1$ and $n_2$ spend $pw$ units each to listen in the first step. 
In the next step node $n_0$ is idle timed to be idle, node $n_1$ spends $pw$ units to transmit to node $n_2$, and $n_2$ spends $pw$ units to listen simultaneously in the second step. So, totally they spend $3pw + 2pw = 5 pw$ units of power to complete the transmission. 
Whereas with the non-timed protocol that uses CSMA, the node $n_0$ spends $pw$ units to transmit to node $n_1$, and simultaneously $n_1$ and $n_2$ spend $pw$ units each to listen in the first step. In the next step nodes $n_0$ and $n_1$ spend $pw$ units each in transmitting but $n_2$ is allowed to listen only to $n_1$, and $n_2$ spends $pw$ units to listen in the second step. So, totally they spend $3pw + 3pw = 6pw$ units of power to complete the exact same transmission. Thus the synthesized protocol uses power efficiently. Also, observe that the non-timed protocol would require CSMA collision detection operations to avoid packet loss which would require use of more (power) resources which are saved in using the synthesized protocol.

\label{sec:Future Work}
\section{Future Work}


One of the shortcomings of expressing requirements in Temporal Logic is that it cannot be used for specifying complex requirements. For example if one of the network requirements was to have loss rate with normal distribution with mean of 20\%, then it becomes impossible to specify such a requirement with temporal logic in the first place. 
We can overcome this problem by using a combination of model repair techniques and genetic programming.
The idea is to use temporal logic to specify only subsets of the
requirements and to generate several individualized protocols
that satisfy these subsets of requirements. The resulting set
of protocols then form a initial population for Genetic Programming
(GP), which eventually through evolution guided by
fitness function (parameterized with missed-out or needed requirements)
provides a solution that satisfies all requirements. More specifically as extended research we plan to do the following,

\textbf{i. Utilizing model repair to generate desired solutions.}

In this project we focused on synthesizing the desired protocol with given specification. It is straightforward that the given specification does not have to be the complete specification for the problem at hand. Instead, we can focus on utilizing only a subset of the specification while identifying constraints for
synthesis. For example, one could only focus on the latency and ignore energy usage. If we pursue this approach, then the synthesized program would only satisfy the latency requirements but not energy usage. We can, thereafter, revise the given program to add the missing properties using the approach of model repair.
Model repair focuses on revising an existing program to satisfy a new property while preserving existing properties.

\textbf{ii. Synthesizing programs that satisfies all requirements using GP.}

A disadvantage of Genetic Programming is that identifying the initial population of possible protocols is tricky. In particular, if the initial population consists of arbitrary programs, then they may utterly fail to design the desired objective. In this case, the success rate of GP can be very low (even 0 in some cases). 
However, we plan to overcome this issue in our future research work by using the set of protocols obtained from the approach presented in this project work as the initial population for GP. Further we will use fitness functions parameterized with missed-out or needed complex requirements and eventually through evolution guided by this fitness function one can reach a solution that satisfies all requirements. 

\label{sec:Summary and Lessons Learned}
\section{Summary and Lessons Learned}


Thus in this project we presented an approach for automated synthesis of network protocols from network specifications or requirements. We followed a seven step framework where for a given communication network we first identify and specify the network requirements in natural language, then specify each of these requirements formally in temporal logic. The formal specifications are then expressed as assertions in Z3 using the Z3Py API. The Z3 SMT solver then checks if these assertions or constraints are satisfiable and returns a model or counterexamples if they are satisfiable or unsatisfiable respectively. The acquired model is the required network protocol which is further parsed to extract and interpret the detailed protocol traces. 

Using this synthesis approach we synthesized two protocols one that performed broadcasting and the other performed data transmission in a restricted neighbor environment. These protocols were simulated in NS3 network simulator to evaluate their applicability. Since the requirements were based on timed reasoning the protocols suggested timed transmissions which were only capable of transmitting less data when compared to non-timed protocols. We used existing non-timed wireless protocol transmission among wifi nodes with CSMA capability as benchmark. Though the benchmark protocol performed better in terms of amount of data transmitted in a given duration of time, based on our analysis we identified that they were not energy efficient since nodes performed continuous transmission even when there were no corresponding receiving nodes. On the contrary nodes, with the synthesized protocol, nodes where timed to perform actions like transmit and listen so they were in idle mode whenever they were not timed to perform any required action which saved power.

Some of the advantages of the synthesis approach presented in this report are as follows:

i. Constraints in LTL are closer to natural language so they can be interpreted easily.

ii. Model generation performed by the Z3 is very fast if the constraints are encoded efficiently and sing Z3Py API helps in encoding constraints efficiently.

However there are also some difficulties in using the presented synthesis approach which are as follows:

i. The Z3 generated model is not easily accessible for further use, for instance it is not possible to answer questions like "what happens at process $x$ at time $y$ as per the model?". This requires detailed multi-level parsing of the stored model-file.

ii. Debugging constraints is tedious due to the non-determinism involved. This is especially the case when we have multiple nested quantification involved in the constraints which are unavoidable when it comes to expressing requirements which involve concurrency. Nested quantification results in lengthy constraints and specifying valid domains for constants in the uninterpreted functions in such constraints are easily prone to errors and are hard to debug.

During the course of this project I learnt that setting goals that can be accomplished is an important part of planning and working on any project. I also learnt that trying to learn too many new things at a the same time can affect the quantity and quality of work. I was new to Z3 at the beginning of this project which made it very hard for me to debug and traceback to specific constants or variables or uninterpreted functions that was causing unexpected behavior in the resulting model or in the the constraints being declared as "unsat". With time when I was finally comfortable with Z3, I had to simulate my protocols in NS3 which was also new to me and the time available was very little for me to  learn concepts in NS3 and the to simulate my protocol from scratch. So I had to customize existing protocols to meet my needs. But I look forward to implementing them from scratch as soon as get comfortable in the NS3 environment. I would recommend future students to take up simpler projects or larger projects only if they already have some basic knowledge in the relevant topic so that they can achieve results and perform evaluation as early as possible, so that they have time to be able to update their methods or consider alternative ways of evaluation if needed.

\ifCLASSOPTIONcaptionsoff
  \newpage
\fi

\bibliography{vidhya}
\bibliographystyle{ieeetran}



\end{document}